# Metal-insulator Transition in a Pyrochlore-type Ruthenium oxide, $Hg_2Ru_2O_7$

Ayako Yamamoto[1,2,*], Peter A. Sharma[1], Yoshihiko Okamoto[1,2], Aiko Nakao[1], Hiroko Aruga Katori[1,2], Seiji Niitaka[1,2], Daisuke Hashizume[1], and Hidenori Takagi[1,2,3]

[1] *RIKEN (The Institute of Physical and Chemical Research), Hirosawa Wako, 351-0198*

[2] *CREST, Japan Science and Technology (JST),*

[3] *Dept. Advanced Materials, Univ. Tokyo, Kashiwa-no-ha, Kashiwa 277-8561*

A new pyrochlore ruthenium oxide, $Hg_2Ru_2O_7$, was synthesized under a high pressure of 6 GPa. In contrast to the extensively studied $Ru^{4+}$ oxides, this compound possesses a novel $Ru^{5+}$ valence state, corresponding to a half-filled $t_{2g}^3$ electron configuration. $Hg_2Ru_2O_7$ exhibits a first order metal-insulator transition at 107 K, accompanied by a structural transition from a cubic to lower symmetry. The behavior of the magnetic susceptibility suggests the possible formation of a spin singlet in the insulating low temperature state.

* Corresponding author e-mail: ayako-yamamoto@postman.riken.jp

Transition metal oxides with the pyrochlore structure (of the formula $A_2B_2O_7$) that possess magnetic 4$d$- and 5$d$-ions in the $B$-site often reveal an interesting interplay between strong electron correlation and magnetism[1]. The relatively wide bandwidth of 4$d$- and 5$d$-systems, compared with 3$d$-transition metal oxides, renders them metallic at room temperature more frequently[2]. Nevertheless, correlations among 4$d$- and 5$d$-electrons still play an important role and often bring about a metal-insulator transition (MIT) as a function of temperature ($T$)[3,4] and chemical substitution[5]. The presence of geometrical frustration on the network of corner sharing tetrahedra in the $B$ ion-sublattice makes the pyrochlore structure particularly interesting to study from the point of view of local moment magnetism. The pyrochlore lattice is known to give rise to one of the strongest degrees of frustration and prevents the long range ordering of spins, charge, and perhaps orbitals. Indeed, we may anticipate that frustration modifies the manner in which the MIT occurs and gives rise to a novel self-organization of a spin-charge-orbital complex.

Among a wide variety of 4$d$- and 5$d$-pyrochlore oxides, pyrochlores with tetravalent ruthenium $Ru^{4+}$ occupying the B-site, $A^{3+}_2Ru^{4+}_2O_7$, have been one of the most intensively studied[4-6]. $Ru^{4+}$ is in the low spin state with four electrons in its nearly degenerate $t_{2g}$ orbitals. The wide variety of ground states in $Ru^{4+}$ pyrochlores is likely due to their proximity to a MIT and the fact that their bandwidths depend strongly on the ionic radius of the trivalent $A^{3+}$ ions through Ru-O-Ru bond bending[7]. $RE^{3+}_2Ru^{4+}_2O_7$ (where $RE$ is a rare earth ion, $e.g.$, Y[5]) is known to be a Mott insulator with a spin-glass ground state. $Bi^{3+}_2Ru^{4+}_2O_7$ remains a metal down to the lowest measured temperatures[6]. $Tl^{3+}_2Ru^{4+}_2O_7$ is a metal at room temperature but exhibits a metal to spin-singlet insulator transition with decreasing $T$[8,9]. A novel interplay between orbital ordering and the $S = 1$ spins of $Ru^{4+}$ has been suggested as the physics behind the spin-singlet formation.

In contrast to $Ru^{4+}$ pyrochlores, only two $Ru^{5+}$ pyrochlores, $Cd_2Ru_2O_7$[10] and $Ca_2Ru_2O_7$[11]), have been reported to date, partly due to the difficulty in oxidizing Ru to be pentavalent. $Ru^{5+}$ has three $t_{2g}$ electrons such as $Cr^{3+}$ and $Mn^{4+}$ and should give rise to a $S = 3/2$ spin moment without orbital degrees of freedom in the ionic limit. In terms of orbital physics, $Ru^{5+}$ pyrochlore should thus be drastically different from the $Ru^{4+}$ pyochlores. The $Ru^{5+}$ pyrochlore might also be viewed as an "itinerant" analog of the geometrically frustrated magnets $ZnCr_2O_4$[12]) and $CdCr_2O_4$[13]). These two chromates are known to be very good insulators, and although a spin liquid state is robust down to low $T$ (~ 10 K), they eventually show transitions to a $S = 3/2$ antiferromagnetically ordered state accompanied by a lattice distortion. The properties of $Cd_2Ru_2O_7$ and $Ca_2Ru_2O_7$, have not yet been completely explored. However, it was reported that both have a relatively low resistivity ($\rho$) of the order of m$\Omega$ cm but do not show a metallic $T$ dependence (d$\rho$/d$T$ > 0) [10,11].

Here, we report a new pyrochlore oxide $Hg_2Ru_2O_7$ as the third member of the $Ru^{5+}$ pyrochlores. In contrast to $Cd_2Ru_2O_7$ and $Ca_2Ru_2O_7$, we found a well-defined MIT at ~ 107 K in $Hg_2Ru_2O_7$. From the magnetic susceptibility ($\chi$) data, we suggest that the ground state of this system is likely to be a spin-singlet state. A close similarity to $Tl_2Ru_2O_7$ strongly suggests that spin singlet formation is not a consequence of a specific orbital configuration but that it may be quite a common occurrence in such transition metal oxides.

Polycrystalline samples were prepared from a stoichiometric mixture of HgO (99.9%) and $RuO_2$ (99.99%). The pressed powder including the oxidizer $KClO_3$ in a gold cell was sintered at 800–950 °C for 1–2 h under 6 GPa using a cubic-anvil-type high-temperature and high-pressure apparatus (TRY Co.). Both the suppression of HgO decomposition and the stabilization of $Ru^{5+}$ under high oxygen pressure underline the necessity and importance of high pressure synthesis. Powder X-ray diffraction (XRD) patterns of the sample obtained by monochromatic Cu $K\alpha_1$ radiation (MacScience, MXP18AHF) at room temperature confirmed a single phase of a cubic pyrochlore-type structure (space group *Fd-3m*, No.227). The structure was refined by the Rietveld method using the program RIETAN2000[14]). All atomic positions assumed the regular position of the pyrochlore structure except for $x(O_1)$ in the 48f site. Occupancy in all atoms was fixed at 1.0, and $B(O_1)$ and $B(O_2)$ fixed at 0.6 Å$^2$ and 1.0 Å$^2$, respectively. The refined structure parameters are as follows: $a$ = 10.199(1) Å, $x(O_1)$ = 0.3182(4), $B(Hg)$ = 0.861(8) Å$^2$, and $B(Ru)$ = 0.554(11) Å$^2$. The reliability factors are $R_{wp}$ = 10.95 %, $R_e$ = 8.08 %, $S$ = 1.35, $R_I$ = 2.90 %, and $R_F$ = 2.49%. The $T$ dependence of XRD was measured using conventional CuKα radiation (MacScience, MXP21TA). The lattice parameter below 300 K was determined by the least-squares method. The metal ratio determined by an electron microprobe analyzer (JEOL) was Hg/Ru = 1.01(2). The chemical analysis of the oxygen content could not be carried out in this study since it was very difficult to dissolve the sample in any acid. KCl remained in the sample as a byproduct of KClO$_3$ and was washed with water as necessary.

X-ray photoemission spectra (XPS) were measured using a VG ESCALAB 250 spectrometer (Thermo-Fisher Electron Co.) employing monochromatic Al-K radiation at room temperature. The spectra of the peaks of Ru 3$d$ were recorded with an energy step of 0.1 eV. The sample was mounted on Au disks as a sample holder that is also used for the standardization of Au 4$f_{7/2}$ (83.9 eV). XPS data of Ru3d for Hg$_2$Ru$_2$O$_7$ and RuO$_2$ are shown in Fig. 1. The binding energy of the 3$d_{5/2}$ peak of Hg$_2$Ru$_2$O$_7$ (281.3 eV) is located 0.6 eV below that of RuO$_2$ (280.7 eV) with Ru$^{4+}$. This clearly demonstrates that a novel valence of Ru$^{5+}$ ($t_{2g}^3$) is stabilized in this compound.

The electrical resistivity, thermoelectric power, heat capacity, and thermal conductivity were measured in a Physical Property Measurement System (Quantum Design). The dc-susceptibility was measured with a Magnetic Property Measurement System (Quantum Design).

A MIT is observed in the $\rho(T)$ of Hg$_2$Ru$_2$O$_7$ at $T_{MI}$ ~ 107 K, as shown in Fig. 2. The $\rho(T)$ curve shows a metallic behavior (d$\rho$/d$T$ > 0) for $T$ > $T_{MI}$. The magnitude of $\rho(T)$ at 300 K was as high as 10 mΩ cm. Considering that the sample is a porous ceramic, its intrinsic resistivity should be substantially lower. Undoubtedly, $Hg_2Ru_2O_7$ is a metal above ~107 K. The Seebeck coefficient $S$ shown in Fig. 2, which is known to be insensitive to grain boundaries, is positive; more importantly, it shows a nearly $T$-linear, metallic $T$-dependence and is as small as 6 μV/K at 300 K typical of conventional metals.

The $\chi(T)$ data shown in Fig. 2 may indicate that the metallic state above ~107 K has a large electronic density of states at the Fermi level ($N(E_F)$). Above ~107 K, we observe a weakly $T$ dependent paramagnetic $\chi$. A Curie-Weiss (CW) fit in the range ~150–300 K yielded a large $\Theta_{CW}$ ~ –2700 K with an effective moment $\mu_{eff}$ ~3.7 $\mu_B$, which is close to that expected for localized $Ru^{5+}$ $S = 3/2$ spins (3.873 $\mu_B$). This value of $\Theta_{CW}$ appears to be uncharacteristically large in comparison with insulating pyrochlores, where $\Theta_{CW}$ often ranges from ~–500 K to –1100 K ($Y_2Ru_2O_7$[5]). We note that a large, weakly $T$-dependent Pauli contribution to $\chi(T)$ also easily explains the data. Given the rather unphysical value of $\Theta_{CW}$, we believe the paramagnetic $\chi(T)$ above 107 K may originate from conduction electron Pauli paramagnetism. A room temperature value of $\chi$ ~ 0.95×10$^{-3}$ emu/mole Oe implies an $N(E_F)$ ~ 14.7 /eV/Ru atom, which corresponds to a $T$-linear specific heat ($C$) $\gamma$ coefficient of ~ 22 mJ/mol K$^2$. This estimation of $\gamma$ is substantially larger than those of conventional metals, strongly suggesting the presence of a relatively narrow 4$d$ band and thus also electron correlations in this system. Interestingly, the XPS data also indicate the presence of strong correlation through the appearance of screened and unscreened peaks in the Ru3$d$ core level spectra[15].

The magnitude of $\rho(T)$ jumps by about one order of magnitude at ~107 K, followed by an insulating $T$-dependence ($d\rho/dT < 0$). We observe a very clear anomaly in $\chi$, $S$, and $C_p$ at the same $T$ shown in Fig. 2. Since $\rho(T)$ and $\chi(T)$ were accompanied by a small hysteresis, the MIT is definitely of first order.

The low $T$ insulating phase shows a nearly $T$-independent, significantly reduced $\chi$, much lower value than that in the metallic state, as shown in Fig. 2. Given the presence of strong electron correlations inferred from the enhanced $\gamma$, it may be natural to suggest that the low $T$ insulating phase is a sort of Mott insulator. Charge density wave formation (CDW) is rather difficult to imagine in this three-dimensional system. The formation of the Mott-insulating state is often accompaniedly unusual magnetism such that we anticipate some sort of $S = 3/2$ pyrochlore magnet due to the $t_{2g}^3$ configuration of $Ru^{5+}$.

The strong geometrical frustration inherent in the pyrochlore structure may lead to an $S = 3/2$ spin liquid followed by a marginally achieved ordered state, as seen in the $S = 3/2$ spinel antiferromagnets (with pyrochlore magnetic sublattices) $ZnCr_2O_4$[12] and $CdCr_2O_4$[13]. However, the $\chi(T)$ observed in $Hg_2Ru_2O_7$ is more than one order of magnitude smaller than that of the $S = 3/2$ pyrochlore antiferromagnets. This paramagnetic $\chi$ is comparable to that of a spin-singlet insulator such as $MgTi_2O_4$ ($S = 1/2$) [16] and $Tl_2Ru_2O_7$ ($S = 1$) [11]. These comparisons lead us to suggest that the $S = 3/2$ moment is quenched into a spin-singlet insulator ground state in $Hg_2Ru_2O_7$.

A structural distortion associated with the MIT is inferred from the XRD pattern and the $T$ dependence of the thermal conductivity ($\kappa$). As shown in Fig. 3(a), $\kappa(T)$, a kink is observed at the MIT. Judging from the Weidemann-Franz law, $\kappa$ is dominated by phonon heat transport such that the kink observed at $T_{MI}$ is consistent with a structural transition. The cubic $a$ lattice parameter discontinuously increases at the MIT. Clearly, the lattice expands from the metallic phase to the insulating phase, which is consistent with the importance of electronic correlations. Importantly, the 444 and 880 peaks in the simple cubic unit cell split below $T_{MI}$, indicating the lowering of the crystal symmetry. Our preliminary electron diffraction measurements at low $T$ have indicated the appearance of superlattice spots in addition to those seen in the cubic phase. These observations suggest the presence of a complicated atomic displacement, which might stabilize the spin-singlet state suggested by $\chi(T)$.

The behavior of the $t_{2g}^3$ $Hg_2Ru_2O_7$ system has a striking parallel with that of the $t_{2g}^4$

system $Tl_2Ru_2O_7$. $Tl_2Ru_2O_7$ displays a MIT at ~125 K, accompanied by a structural phase transition from cubic to orthorhombic[4]. It has been argued that the $Tl_2Ru_2O_7$ orbital ordering in the insulating state gives rise to an essentially one-dimensional chain of Ru $S = 1$ antiferromagnetically coupled moments[9]. A one-dimensional chain of integer spins would then be unstable towards the formation of a valence bond singlet state, associated with the appearance of the Haldane gap. $Hg_2Ru_2O_7$, with its $t_{2g}^3$ configuration, should have $S = 3/2$ spins without any orbital degree of freedom. The spin gap formation of half-integer spins is not uncommon in materials without quasi-one-dimensional structures. The similarity between the two system may suggest that the spin singlet formation is quite a common phenomena of correlated pyrochlores and not necessarily the consequence of $S = 1$ quantum magnetism.

A comparison of the other $Ru^{5+}$ pyrochlores, $Cd_2Ru_2O_7$[10] and $Ca_2Ru_2O_7$[11], is also rather informative, though the physical and structural properties of those two compounds have yet to be clarified. These two pyrochlores do not show any clear evidence for a MIT as in $Hg_2Ru_2O_7$, but $\rho(T)$ weakly increases with decreasing $T$; this suggests a close proximity to an insulator. The presence of frustrated magnetic moments is inferred from the spin-glass behavior observed in $Ca_2Ru_2O_7$[10], indicative of the importance of electron correlations. These observations appear to imply that all these pyrochlores are commonly located in the vicinity of a correlation driven MIT and that a subtle difference in local structure, hybridization with $A^{2+}$ ions, disorde,r and so on brings these systems to distinctly different ground states. The unit cell parameter, $a$, of $Hg_2Ru_2O_7$ ($a = 10.199(1)$ Å) is comparable to that of $Ca_2Ru_2O_7$ ($a = 10.197(2)$ Å[11]) and distinctly larger than that of $Cd_2Ru_2O_7$ ($a = 10.1291$) Å[10]). It is rather difficult to find a definite correlation between the unit cell parameter, *i.e.* Ru–Ru distance, and the ground state.

In conclusion, we found a first-order metal-insulator transition at 107 K accompanied by a structural change in the new pyrochlore $Hg_2Ru_2O_7$, prepared under high pressure. The novel $Ru^{5+}$ valence in this compound was confirmed by XPS. The insulating state is likely to be a spin singlet; however, future studies are needed in order to confirm this. $Hg_2Ru_2O_7$ provides us an intriguing topic for the study of metal-insulator transitions in the presence of strong geometrical frustration.

Note: Just before the submission of this paper, we became aware of a parallel study on $Hg_2Ru_2O_7$ that has been reported by Klein *et al*[17]. Their crystal structure at room temperature and basic physical properties are consistent with our results. However, our sample showed a sharper M-I transition, and this enables us to observe thermal hysteresis and a structural change, which are evidences of a first-order transition.

Acknowledgements

We thank K. Watanabe from the advanced development and supporting center (RIKEN) for the measurements of EPMA. We also thank J. Matsuno, A. Chainani, and M. Uchida for useful discussion.  This work was partly supported by a Grant-in-Aid for scientific research (No. 16GS50219) from the ministry of Education, Culture, Sports, Science and Technology as well as by the RIKEN DRI Research Grant.

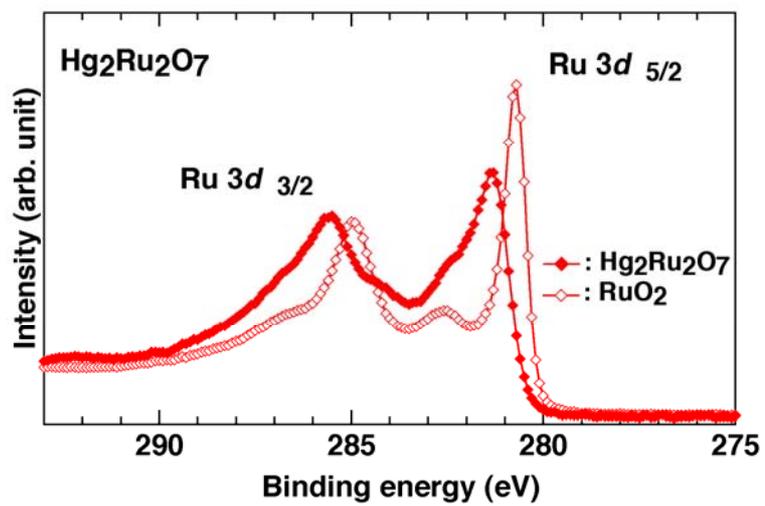

Fig. 1 (Color online) X-ray photoemission spectra of Ru3$d$ for $Hg_2Ru_2O_7$ and $RuO_2$ at room temperature.

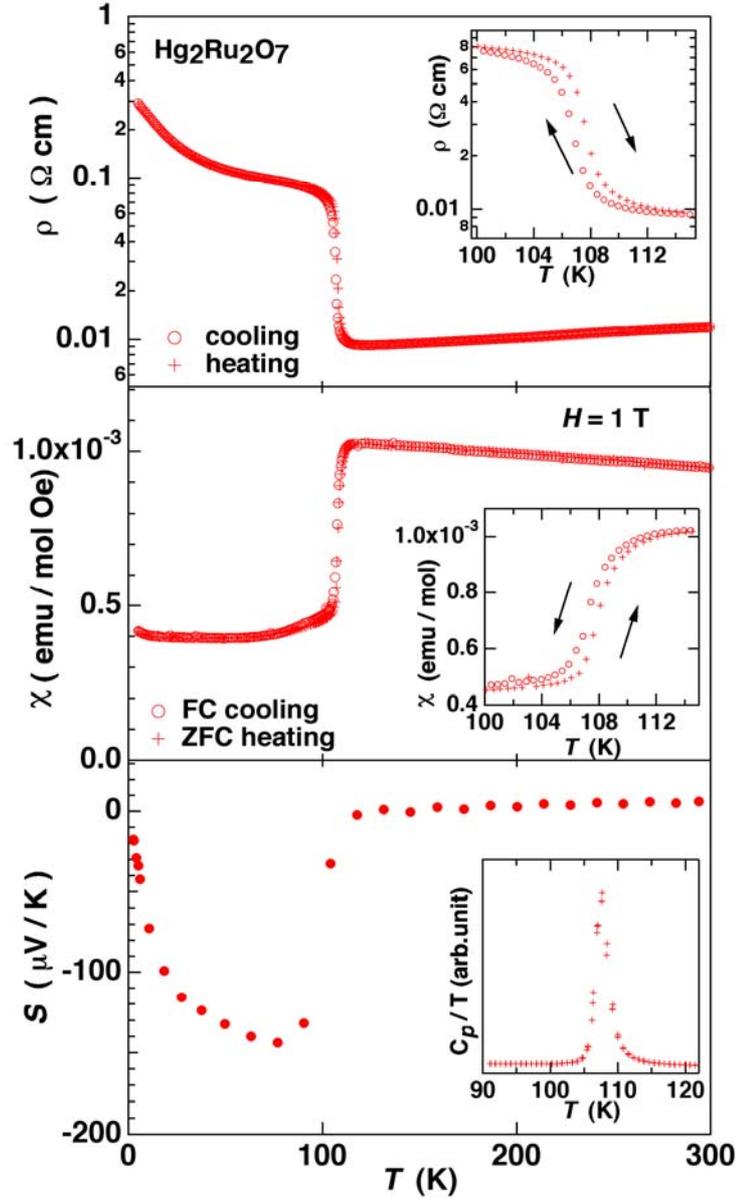

Fig. 2 (Color online) Temperature dependence of electrical resistivity (top), dc-susceptibility (center), and thermoelectric power (bottom) for $Hg_2Ru_2O_7$. The insets in the top and center panels display an enlarged regions showing the hysteresis associated with a first order transition. The inset in the bottom panel displays the specific heat peak observed at the MIT. In the center panel, a small amount of impurity contribution (~5% of the signal) was subtracted from the susceptibility data.

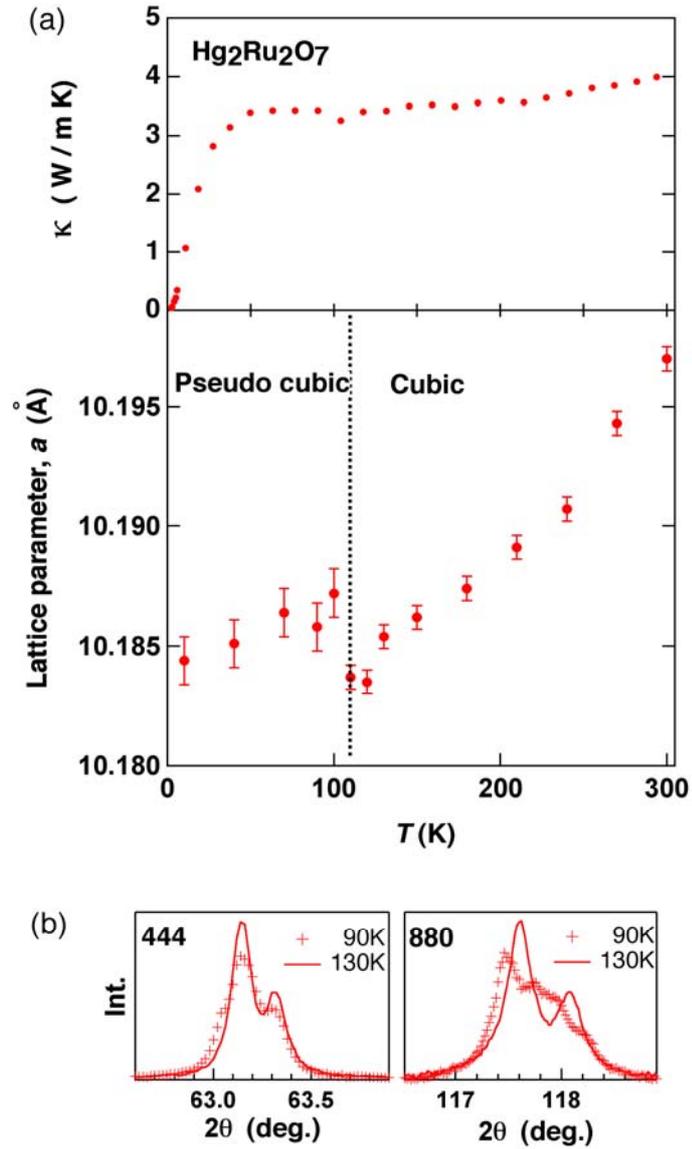

Fig. 3 (Color online)  (a) Temperature dependence of thermal conductivity and lattice parameter, *a* for $Hg_2Ru_2O_7$. The lattice parameter at both high and low temperatures was tentatively indexed with a cubic cell. The dotted line is a guide for the eyes. (b) X-ray powder diffraction patterns (Cu $K\alpha_1$ and $\alpha_2$) above (130 K) and below (90 K) the transition temperature for the selected indexes. Note that the peak shape changes are probably due to a lowering of symmetry.